\documentstyle[12pt]{article} 
\begin{document}

\overfullrule 0 mm
\language 0
\centerline { \bf{ CAN MECHANICALLY EXTENDED SOURCES OF RADIATION}}
\centerline { \bf{IN CLASSICAL ELECTRODYNAMICS SOLVE ALL PROBLEMS}}
\centerline { \bf{ OF RADIATING POINT-LIKE SOURCES?}}
 \vskip 0.5 cm
\centerline {\bf{ Alexander A.  Vlasov}}
\vskip 0.3 cm
\centerline {{  High Energy and Quantum
Theory}} \centerline {{  Department of Physics}} \centerline {{
Moscow State University}} \centerline {{  Moscow, 119899}}
\centerline {{  Russia}}
\vskip 0.3 cm
{\it On simple example is shown that a model of mechanically extended
source of radiation in classical electrodynamics indeed has
some problems with radiation reaction force. So one cannot, in
general case, state, that the present situation with classical
dynamics of mechanically extended sources of radiation is "very
satisfactory".}

03.50.De
\vskip 0.3 cm
In classical electrodynamics there are known problems of radiation
reaction force for point-like sources of radiation: preacceleration,
runaway and other pathological (unphysical) solutions.

In the literature one can find the opinion that if the finite size
of a radiating source is fully taken into account all the above
problems can be solved.

 In this paper we examine the well known model of
"nonrelativistically rigid charged sphere" to show that
nevertheless there remain problems with radiation reaction force.

It should be mentioned that "extended (mechanically) electron
models" were proposed long time ago in the works of Abraham, Lorentz,
Sommerfeld and others ( for good reviews see, for ex., [1,2,3,4]).

Thus in Sommerfeld works [5] was derived the expression of self-force
acting on "nonrelativistically rigid charged sphere", i.e sphere with
radius $a$, its center moving along trajectory
$\vec R(t)$, with total charge $Q$ and charge density (in
laboratory reference frame) $$\rho(t,\vec r)={Q\over 4\pi
a^2}\delta(|\vec r- \vec R|-a)$$.

In the case of
shell rectilinear motion in [7] was given for the
self-force
$$\vec F_{self} = \int dV  \rho \left(\vec
E +[\vec v, \vec H]/c \right) $$
the following expression:
 $$F_{ self} = {Q^2c\over
4a}\int\limits_{-1}^{+1}dx\int\limits_{t_1}^{t}dt'{ax+L'\over
 [a^2+2axL'+(L')^2]^{3/2}}$$
$$ +{Q^2\over
4a}\int\limits_{-1}^{+1}dx
{x-v/c \over a(1-xv/c)}  $$
$$-{Q^2\over
4a}\int\limits_{-1}^{+1}dx
{ N_1 - v_1/c \over [a^2+2axL_1+(L_1)^2]^{1/2}
(1- N_1  v_1/c)}  \eqno(1)$$
here
$$L'=R(t)-R(t'),\ \ L_1=R(t)-R(t_1), \ \ v= dR(t)/dt,\ \ v_1=v(t_1),
$$ $$N_1={ax+L_1\over [a^2+2axL_1+(L_1)^2]^{1/2}}  $$ and $t_1$ is
the solution of the equation (equation of retardation):  $$ t_1=t-
[a^2+2axL_1+(L_1)^2]^{1/2}/c-a/c,$$

After integration by parts, expression (1) can be put in the form
$$F_{self}= {Q^2\over 4a^2}\left( {1-\beta^2 \over \beta^2}\ln {1+\beta
\over1-\beta} - \int\limits_{L_{-}}^{L_{+}} {dL_1\over L_1}
{1-\beta^2(t_1(L_1)) \over \beta^2(t_1(L_1))} \right)$$
$$+{Q^2c\over
4a}\int\limits_{-1}^{+1}dx\int\limits_{t_1}^{t}dt'{\dot v'\over
 v'^2 [a^2+2axL'+L'^2]^{1/2}} \eqno(2)$$

here
$$ L_{\pm}=L_{1}(x=\pm 1),\ \ v'=v(t'), \ \ \beta=v/c$$
With (2) equation of shell motion is
$$m{d \over dt}(\gamma v)=F_{self} +F_{ext} \eqno(3)$$
here $m$ is the total mass of the shell, $F_{ext}$ - some external
force.

Let us mention some features of the eq.(3).

1). If $F_{ext}=0$, then eq. (3) has the trivial solution
$$v=v_{0}=const$$
because the expression (1,2) for the self-force identically vanishes.

2). But this solution is not unique. An example of so called
 radiationless motion of a rigid sphere was given in [6].

Now we consider small deviations from  shell uniform rectilinear
motion for linear analysis of stability of such motion.

For this goal it will be convenient to use for the self-force the
Sommerfeld expression [2], which is equivalent to (1)
(this can be shown spreading with the help of step functions
limits of integration over $t'$ in the  double intergal in (1) to
the whole axis $(-\infty,+\infty)$ and changing  the order of
integration over $x$ and $t'$ ):
$$F_{self}={Q^2 \over 4 a^2}\left[ -c \int\limits_{T^{-}}^{T^{+}} dT
{cT-2a \over L^2} + \ln {{L^{+}\over L^{-}}} + ({1\over
\beta^2}-1)\ln { {1+\beta \over 1-\beta}} -{2\over \beta} \right]
\eqno(4)$$ here $cT^{\pm}=2a \pm L^{\pm},\ \
L^{\pm}=|R(t)-R(t-T^{\pm})|,\ \ L=|R(t)-R(t-T)| $.

Let us take
$$R(t)=R_0+vt+\delta(t),\ \ R(t-T)=R_0+v(t-T)+\delta(t-T),\ \
L=vT+\delta(t)-\delta(t-T),$$
here $\delta(t)$ - small perturbation , $v\not=c, \not=0$.

Then in linear approximation the self-force (4) reads
 $$F_{self}={Q^2 \over 4 a^2}\left[
{4\dot{\delta}\over \beta^2}\left(1-{1\over 2\beta}\ln{{(1+\beta)
\over(1-\beta)}}\right)-{2c\over v^3}\int\limits_{T^{-}}^{T^{+}}dT
{cT-2a\over T^3}\delta(t-T)\right] \eqno(5)$$
here $T^{\pm}={2a\over c(1\mp \beta)},$
and the equation of shell motion is
$$m\gamma^3\ddot{\delta}= F_{self} \eqno(6)$$
Solution of eq. (6) we take as $\delta \sim
\exp{(pt)}$.  Then the parameter $p$ must obey the
following equation $$p^2={Q^2\over
a^2m\gamma^3}\left[ {p\over c\beta^2} \left(1-{1\over
2\beta}\ln{{(1+\beta) \over(1-\beta)}}\right) -{c\over
2v^3}\int\limits_{T^{-}}^{T^{+}}dT {cT-2a\over T^3}\exp{(-pT)})\right]
\eqno(7)$$

Now consider two boundary values of shell velocity.

For analysis of stability of the solution $v_0=0$ we examine the
approximation of small velocities. This yields for the self-force
known expression [2]:
$$F_{self}={Q^2\over 3ca^2} \left[ v(t-2a/c) -v(t) \right]
\eqno(8)$$ and the eq. of motion $$m\dot{v} = {Q^2\over 3ca^2}
\left[ v(t-2a/c) -v(t) \right] \eqno(9)$$ does not possess growing
with time solutions.

For analysis of stability of the solution  $v\to c$ it will be
convenient to use the expression (2) for the self-force.
Let's put in (2) $R(t)=R_0+ct+\delta(t)$, then the main
contribution to the eq. of motion gives the first term in (2) and
$$F_{self} \sim \dot{\delta}\ln{|\dot{\delta}|},$$ as the L.H.S. of
eq. of motion takes the form $\sim
\ddot{\delta}\left(-\dot{\delta}\right)^{-3/2},$ then full
eq.  of motion is $$\ddot{\delta}=
-k\left(-\dot{\delta}\right)^{+5/2}\ln{|\dot{\delta}|}$$ here $k$
is positive parameter.

This eq. has the solution $$\delta \sim t^{1/3} \ln^{-2/3}{t},\ \ t
\to \infty $$ that is the uniform motion of the sphere with
 $v\to c$ is relatively stable in that sense that $$\dot{\delta}/c
\to 0,\ \ \delta/(R=ct)\to 0,\ \ t\to\infty \eqno(10).$$

 Thus there can be solutions asymptotically tending to light cone.

 More general analysis of eq.(7) will be done in the next preprint
(under preparation).

3). There is one more problem, actual to all intergal equations.
If the small perturbation $\delta(t)$ is zero for $t<0$, then the
Laplace transformation of (5,6) yields the  equation for eigenvalues
$\lambda$ of this integral eq. equal to the eq. for $p$  (7).

Thus  at least for $\lambda a/c\ll 1$  this eq.  is solvable and
eigenvalue is not zero.

Then, as the theory of intergal equations tells, not for all forms of
external forces the solution  of the inhomogeneous
equation (3) with $F_{ext}\not= 0$  exists!

The similar situation occurs for rotating shell - see [8].

Thus our conclusion is:

  a model of
mechanically extended source of radiation in classical
electrodynamics with retardation indeed possesses some  problems with
radiation reaction force.

So
one cannot, in general, state, that the present situation with
classical dynamics of mechanically extended sources of radiation is
"very satisfactory".

And to solve problems of radiation reaction force one must search for
new ideas and approaches.

 \vskip 0.5 cm \centerline {\bf{REFERENCES}}

  \begin{enumerate}
\item T.Erber, Fortschr.Phys., 9, 342 (1961).
\item P.Pearle, in {\it Electromagnetism}, ed. D.Tepliz, Plenum, NY,
1982, p.211.
\item A.D.Yaghjian, {\it Relativistic Dynamics of a Charged Sphere},
 Lecture Notes in Physics, 11, Springer, Berlin, 1992.
\item F.Rohrlich, Am.J.Physics, 65(11), 1051 (1997).
\item A.Sommerfeld, Gottingen Nachrichten, 29 (1904), 363 (1904), 201
  (1905).
\item G.A.Schott, Phil.Mag., 15, 752 (1933).
\item Alexander A.Vlasov, physics/9711024.

\item Alexander A.Vlasov, physics/9801017.
\end{enumerate}

 \end{document}